**Electrically tunable Gilbert damping in van der Waals heterostructures of two-dimensional ferromagnetic metals and ferroelectrics**

Liang Qiu,[1] Zequan Wang,[1] Xiao-Sheng Ni,[1] Dao-Xin Yao[1,2] and Yusheng Hou [1,*]

AFFILIATIONS

[1] Guangdong Provincial Key Laboratory of Magnetoelectric Physics and Devices, State Key Laboratory of Optoelectronic Materials and Technologies, Center for Neutron Science and Technology, School of Physics, Sun Yat-Sen University, Guangzhou, 510275, China

2 International Quantum Academy, Shenzhen 518048, China

ABSTRACT

Tuning the Gilbert damping of ferromagnetic (FM) metals via a nonvolatile way is of importance to exploit and design next-generation novel spintronic devices. Through systematical first-principles calculations, we study the magnetic properties of the van der Waals heterostructure of two-dimensional FM metal $CrTe_2$ and ferroelectric (FE) $In_2Te_3$ monolayers. The ferromagnetism of $CrTe_2$ is maintained in $CrTe_2/In_2Te_3$ and its magnetic easy axis can be switched from in-plane to out-of-plane by reversing the FE polarization of $In_2Te_3$. Excitingly, we find that the Gilbert damping of $CrTe_2$ is tunable when the FE polarization of $In_2Te_3$ is reversed from upward to downward. By analyzing the $k$-dependent contributions to the Gilbert damping, we unravel that such tunability results from the changed intersections between the bands of $CrTe_2$ and Fermi level on the reversal of the FE polarizations of $In_2Te_3$ in $CrTe_2/In_2Te_3$. Our work provides an appealing way to electrically tailor Gilbert dampings of two-dimensional FM metals by contacting them with ferroelectrics.

*Authors to whom correspondence should be addressed:
[Yusheng Hou, houysh@mail.sysu.edu.cn]

Since the atomically thin long-range ferromagnetic (FM) orders at finite temperatures are discovered in CrI$_3$[1] monolayer (ML) and Cr$_2$Ge$_2$Te$_6$[2] bilayer, two-dimensional (2D) van der Waals (vdW) FM materials have attracted intensive attention.[3-5] Up to now, many novel vdW ferromagnets such as Fe$_3$GeTe$_2$,[6] Fe$_5$GeTe$_2$,[7] VSe$_2$[8,9] and MnSe$_2$[10] have been synthesized in experiments. Due to the intrinsic ferromagnetism in these vdW FM materials, it is highly fertile to engineer emergent phenomena through magnetic proximity effect in their heterostructures.[11] For instance, an unprecedented control of the spin and valley pseudospins in WSe$_2$ ML is reported in CrI$_3$/WSe$_2$.[12] By contacting the thin films of three-dimensional topological insulators and graphene with CrI$_3$, high-temperature quantum anomalous Hall effect and vdW spin valves are proposed in CrI$_3$/Bi$_2$Se$_3$/CrI$_3$[13] and CrI$_3$/graphene/CrI$_3$,[14] respectively. On the other hand, the magnetic properties of these vdW FM materials can also be controlled by means of external perturbations such as gating and moiré patterns.[3] In CrI$_3$ bilayer, Huang *et al*. observed a voltage-controlled switching between antiferromagnetic (AFM) and FM states.[15] Via an ionic gate, Deng *et al*. even increased the Curie temperature ($T_C$) of the thin flake of vdW FM metal Fe$_3$GeTe$_2$ to room temperature, which is much higher than its bulk $T_C$.[6] Very recently, Xu *et al*. demonstrated a coexisting FM and AFM state in a twisted bilayer CrI$_3$.[16] These indicate that vdW FM materials are promising platforms to design and implement spintronic devices in the 2D limit.[4,11]

Recently, of great interest is the emergent vdW magnetic material CrTe$_2$ which is a new platform for realizing room-temperature intrinsic ferromagnetism.[17,18] Especially, CrTe$_2$ exhibits greatly tunable magnetism. In the beginning, its ground state is believed to be the nonmagnetic 2$H$ phase,[19] while several later researches suggest that either the FM or AFM 1$T$ phases should be the ground state of CrTe$_2$.[17,18,20-23] Currently, the consensus is that the structural ground state of CrTe$_2$ is the 1$T$ phase. With respect to its magnetic ground state, a first-principles study shows that the FM and AFM ground states in CrTe$_2$ ML depend on its in-plane lattice constants.[24] It is worth noting that the $T_C$ of FM CrTe$_2$ down to the few-layer limit can be higher than 300 K,[18] making it have wide practical application prospects in spintronics.

Building heterostructures of FM and ferroelectric (FE) materials offers an effective way to control nonvolatile magnetism via an electric field. Experimentally, Eerenstein *et al*. presented an electric-field-controlled nonvolatile converse magnetoelectric effect in a multiferroic heterostructure $La_{0.67}Sr_{0.33}MnO_3/BaTiO_3$.[25] Later, Zhang *et al*. reported an electric-field-driven control of nonvolatile magnetization in a heterostructure of FM amorphous alloy $Co_{40}Fe_{40}B_{20}$ and FE $Pb(Mg_{1/3}Nb_{2/3})_{0.7}Ti_{0.3}O_3$.[26] Theoretically, Chen *et al*. demonstrated based on first-principles calculations that the interlayer magnetism of $CrI_3$ bilayer in $CrI_3/In_2Se_3$ is switchable between FM and AFM couplings by the nonvolatile control of the FE polarization direction of $In_2Se_3$.[27] In spite of these interesting findings, using FE substrates to electrically tune the Gilbert damping of ferromagnets, an important factor determining the operation speed of spintronic devices, is rarely investigated in 2D FM/FE vdW heterostructures. Therefore, it is of great importance to explore the possibility of tuning the Gilbert damping in such kind of heterostructures.

In this work, we first demonstrate that the magnetic ground state of 1$T$-phase $CrTe_2$ ML will change from the zigzag AFM (denoted as z-AFM) to FM orders with increasing its in-plane lattice constants. By building a vdW heterostructure of $CrTe_2$ and FE $In_2Te_3$ MLs, we show that the magnetic easy axis of $CrTe_2$ can be tuned from in-plane to out-of-plane by reversing the FE polarization of $In_2Te_3$, although its ferromagnetism is kept. Importantly, we find that the Gilbert damping of $CrTe_2$ is tunable with a wide range on reversing the FE polarization of $In_2Te_3$ from upward to downward. Through looking into the $k$-dependent contributions to the Gilbert damping, we reveal that such tunability originates from the changed intersections between the bands of $CrTe_2$ and Fermi level when the FE polarizations of $In_2Te_3$ is reversed in $CrTe_2/In_2Te_3$. Our work demonstrates that putting 2D vdW FM metals on FE substrates is an attractive method to electrically tune their Gilbert dampings.

$CrTe_2$, a member of the 2D transition metal dichalcogenide family, can potentially crystalize into several different layered structures such as 1$T$, 1$T_d$, 1$H$ and 2$H$ phases.[28] It is believed that the 1$T$ phase is the most stable among all of these possible phases in both bulk and ML. This phase has a hexagonal lattice and belongs to the $P3m1$ space group, with each Cr atom surrounded by the octahedrons of Te atoms (Fig. 1a). In view of the hot debates on the magnetic ground state in CrTe$_2$ ML, we establish a $2\times2\sqrt{3}$ supercell and calculate the total energies of several different magnetic structures (Fig. S1 in Supplementary Materials) when its lattice constant varies from 3.65 to 4.00 Å. As shown in Fig. 1b, our calculations show that z-AFM order is the magnetic ground state when the lattice constant is from 3.65 to 3.80 Å. By contrast, the FM order is the magnetic ground state when the lattice constant is in the range from 3.80 to 4.00 Å. Note that our results are consistent with the experimentally observed z-AFM[23] and FM[29] orders in CrTe$_2$ with a lattice constant of 3.70 and 3.95 Å, respectively. Since we are interested in the Gilbert damping of ferromagnets and the experimentally grown CrTe$_2$ on ZrTe$_2$ has a lattice constant of 3.95 Å,[29] we will focus on CrTe$_2$ ML with this lattice constant hereinafter.

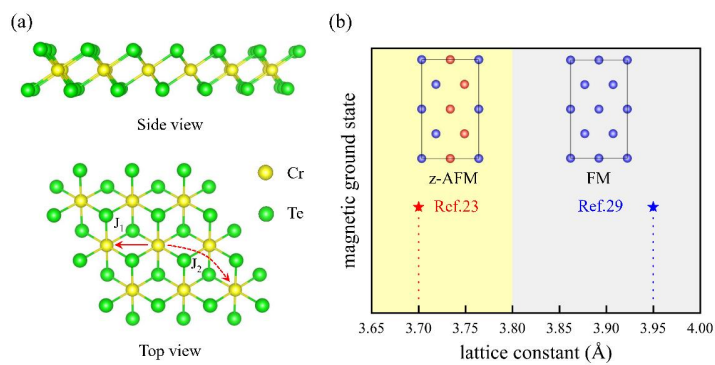

FIG. 1. (a) Side (the top panel) and top (the bottom panel) views of CrTe$_2$ ML. The NN and second-NN exchange paths are shown by red arrows in the top view. (b) The phase diagram of the magnetic ground state of CrTe$_2$ ML with different lattice constants. Insets show the schematic illustrations of the z-AFM and FM orders. The up and down spins are indicated by the blue and red balls, respectively. The stars highlight the experimental lattice constants of CrTe$_2$ in Ref.23 and Ref.29.

To obtain an deeper understanding on the ferromagnetism of CrTe$_2$ ML, we adopt a spin Hamiltonian consisting of Heisenberg exchange couplings and single-ion magnetic anisotropy (SIA) as follows:[30]

$$H = J_1 \sum_{\langle ij \rangle} S_i \cdot S_j + J_2 \sum_{\langle\langle ij \rangle\rangle} S_i \cdot S_j - A \sum_i (S_i^z)^2 \qquad (1)$$

In Eq. (1), $J_1$ and $J_2$ are the nearest neighbor (NN) and second-NN Heisenberg exchange couplings. Note that a negative (positive) $J$ means a FM (AFM) Heisenberg exchange couples. Besides, $A$ parameterizes the SIA term. First of all, our DFT calculations show that the magnetic moment of CrTe$_2$ ML is 3.35 $\mu_B$/Cr, consistent with previous DFT calculations.[31] As shown in Table I, the calculated $J_1$ and $J_2$ are both FM and $J_1$ is much stronger than $J_2$. Both FM $J_1$ and $J_2$ undoubtedly indicate that CrTe$_2$ ML has a FM magnetic ground state. Finally, the SIA parameter $A$ is obtained by calculating the energy difference between two FM states with out-of-plane and in-plane magnetizations. Our calculations obtain $A$=1.81 meV/Cr, indicating that CrTe$_2$ ML has an out-of-plane magnetic easy axis. Hence, our calculations show that CrTe$_2$ ML exhibits an out-of-plane FM order, consistent with experimental observations.[29]

TABLE I. Listed are the in-plane lattice constants $a$, Heisenberg exchange couplings $J$ (in unit of meV) and SIA (in unit of meV/Cr) of CrTe$_2$ ML and CrTe$_2$/In$_2$Te$_3$.

| System | $a$ (Å) | $J_1$ | $J_2$ | $A$ |
| --- | --- | --- | --- | --- |
| CrTe$_2$ | 3.95 | -24.56 | -0.88 | 1.81 |
| CrTe$_2$/In$_2$Te$_3$(↑) | 7.90 | -20.90 | -1.80 | -1.44 |
| CrTe$_2$/In$_2$Te$_3$(↓) | 7.90 | -19.33 | -0.88 | 0.16 |

To achieve an electrically tunable Gilbert damping in CrTe$_2$ ML, we establish its vdW heterostructure with FE In$_2$Te$_3$ ML. In building this heterostructure, we stack a 2×2 supercell of CrTe$_2$ and a $\sqrt{3}\times\sqrt{3}$ supercell of In$_2$Te$_3$ along the (001) direction. Because the magnetic properties of CrTe$_2$ ML are the primary topic and the electronic properties of In$_2$Te$_3$ ML are basically not affected by a strain (Fig. S2), we stretch the lattice constant of the latter to match that of the former. Fig. 2a shows the most stable stacking configuration in CrTe$_2$/In$_2$Te$_3$ with an upward FE polarization [denoted as CrTe$_2$/In$_2$Te$_3$(↑)]. At the interface in this configuration, one of four Cr atoms and one of four Te atoms at the bottom of CrTe$_2$ sits on the top of the top-layer Te atoms of In$_2$Te$_3$. In CrTe$_2$/In$_2$Te$_3$ with a downward FE polarization [denoted as CrTe$_2$/In$_2$Te$_3$(↓)], the stacking configuration at its interface is same as that in CrTe$_2$/In$_2$Te$_3$(↑). The only difference between CrTe$_2$/In$_2$Te$_3$(↑) and CrTe$_2$/In$_2$Te$_3$(↓) is that the middle-layer Te atoms of In$_2$Te$_3$ in the former is farther to CrTe$_2$ than that in the latter (Fig. 2a and 2c). It is noteworthy that the bottom-layer Te atoms of CrTe$_2$ do not stay at a plane anymore in the relaxed CrTe$_2$/In$_2$Te$_3$ (see more details in Fig. S3), suggesting non-negligible interactions between CrTe$_2$ and In$_2$Te$_3$.

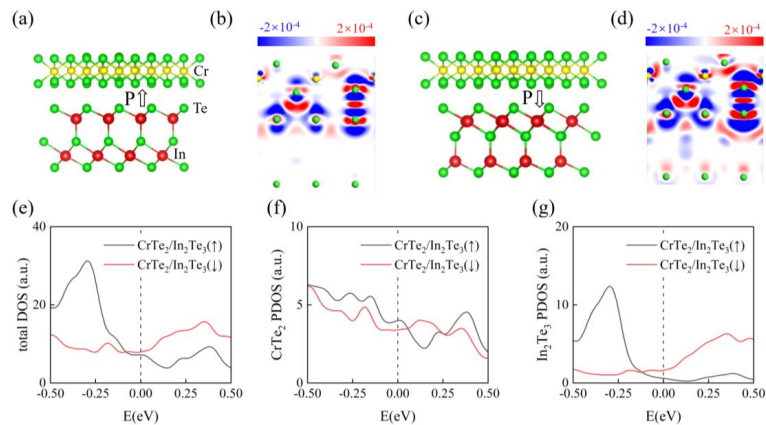

FIG. 2. (a) The schematic stacking configuration and (b) charge density difference $\Delta\rho$ of CrTe$_2$/In$_2$Te$_3$(↑). (c) and (d) same as (a) and (b) but for CrTe$_2$/In$_2$Te$_3$(↓). In (b) and (d), color bar indicates the weight of negative (blue) and positive (red) charge density differences. (e) The total DOS of CrTe$_2$/In$_2$Te$_3$. (f) and (g) show the PDOS of CrTe$_2$ and In$_2$Te$_3$ in CrTe$_2$/In$_2$Te$_3$, respectively. In (e)-(g), upward and downward polarizations are indicated by black and red lines, respectively.

To shed light on the effect of the FE polarization of In$_2$Te$_3$ on the electronic property of CrTe$_2$/In$_2$Te$_3$, we first investigate the spatial distribution of charge density difference $\Delta\rho = \rho_{CrTe_2/In_2Te_3} - \rho_{CrTe_2} - \rho_{In_2Te_3}$ with different FE polarization directions. As shown in Fig. 2b and 2d, we see that there is an obvious charge transfer at the interfaces of both CrTe$_2$/In$_2$Te$_3$(↑) and CrTe$_2$/In$_2$Te$_3$(↓), which is further confirmed by the planar averaged $\Delta\rho$ (Fig. S4). Additionally, the charge transfer in CrTe$_2$/In$_2$Te$_3$(↑) is distinctly less than that in CrTe$_2$/In$_2$Te$_3$(↓). Fig. 2e shows that the total density of states (DOS) near Fermi level are highly different in CrTe$_2$/In$_2$Te$_3$(↑) and CrTe$_2$/In$_2$Te$_3$(↓). By projecting the DOS onto CrTe$_2$ and In$_2$Te$_3$, Fig. 2f shows that the projected DOS (PDOS) of CrTe$_2$ in CrTe$_2$/In$_2$Te$_3$(↑) is larger than that in CrTe$_2$/In$_2$Te$_3$(↓) at Fermi level. Interestingly, the PDOS of In$_2$Te$_3$ in CrTe$_2$/In$_2$Te$_3$(↑) is larger than that in CrTe$_2$/In$_2$Te$_3$(↓) below Fermi level while the situation is inversed above Fermi level (Fig. 2g). By looking into the five Cr-$d$ orbital projected DOS in CrTe$_2$/In$_2$Te$_3$(↑) and CrTe$_2$/In$_2$Te$_3$(↓) (Fig. S5), we see that there are obviously different occupations for $d_{xy}$, $d_{x^2-y^2}$ and $d_{3z^2-r^2}$ orbitals near Fermi level. All of these imply that the reversal of the FE polarization of In$_2$Te$_3$ may have an unignorable influence on the magnetic properites of CrTe$_2$/In$_2$Te$_3$.

Due to the presence of the FE In$_2$Te$_3$, the inversion symmetry is inevitably broken and nonzero Dzyaloshinskii-Moriya interactions (DMIs) may exist in CrTe$_2$/In$_2$Te$_3$. In this case, we add a DMI term into Eq. (1) to investigate the magnetism of CrTe$_2$/In$_2$Te$_3$ and the corresponding spin Hamiltonian is in the form of[32]

$$H = J_1 \sum_{\langle ij \rangle} S_i \cdot S_j + J_2 \sum_{\langle\langle ij \rangle\rangle} S_i \cdot S_j + \sum_{\langle ij \rangle} \boldsymbol{D}_{ij} \cdot (S_i \times S_j) - A \sum_i (S_i^z)^2 \qquad (2).$$

In Eq. (2), $\boldsymbol{D}_{ij}$ is the DMI vector of the NN Cr-Cr pairs. As the $C_6$-rotational symmetry with respect to Cr atoms in CrTe$_2$ is reduced to the $C_3$-rotational symmetry, the NN DMIs are split into four different DMIs (Fig. S6). For simplicity, the $J_1$ and $J_2$ are still regarded to be six-fold. From Table I, we see that the NN $J_1$ of both CrTe$_2$/In$_2$Te$_3$(↑) and CrTe$_2$/In$_2$Te$_3$(↓) are still FM but slightly smaller than that of free-standing CrTe$_2$ ML. Moreover, the second-NN FM $J_2$ is obviously enhanced in CrTe$_2$/In$_2$Te$_3$(↑) compared with CrTe$_2$/In$_2$Te$_3$(↓) and free-standing CrTe$_2$ ML. To calculated the NN DMIs, we build a $\sqrt{3}\times\sqrt{3}$ supercell of CrTe$_2$/In$_2$Te$_3$ and the four-state method[33] is employed here. As listed in Table S1, the FE polarization direction of $In_2Te_3$ basically has no qualitative effect on the DMIs in $CrTe_2/In_2Te_3$ although it affects their magnitudes. More explicitly, the magnitudes of the calculated DMIs range from 1.22 to 2.81 meV, which are about one order smaller than the NN $J_1$. Finally, we find that the SIA of $CrTe_2/In_2Te_3$ is strongly dependent on the FE polarization of $In_2Te_3$. When $In_2Te_3$ has an upward FE polarization, the SIA of $CrTe_2/In_2Te_3(\uparrow)$ is negative, indicating an in-plane magnetic easy axis. However, when the FE polarization of $In_2Te_3$ is downward, $CrTe_2/In_2Te_3(\downarrow)$ has a positive SIA, indicating an out-of-plane magnetic easy axis. It is worth noting that $CrTe_2/In_2Te_3(\downarrow)$ has a much weak SIA than the free-standing $CrTe_2$ ML, although they both have positive SIAs. The different Heisenberg exchange couplings, DMIs and SIAs in $CrTe_2/In_2Te_3(\uparrow)$ and $CrTe_2/In_2Te_3(\downarrow)$ clearly unveil that the magnetic properties of $CrTe_2$ are tuned by the FE polarization of $In_2Te_3$.

To obtain the magnetic ground state of $CrTe_2/In_2Te_3$, MC simulations are carried out. As shown in Fig. S7, $CrTe_2/In_2Te_3(\uparrow)$ has an in-plane FM magnetic ground state whereas $CrTe_2/In_2Te_3(\downarrow)$ has an out-of-plane one. Such magnetic ground states are understandable. Firstly, the ratios between DMIs and the NN Heisenberg exchange couplings are small and most of them are out of the typical range of 0.1–0.2 for the appearance of magnetic skyrmions.[34] Secondly, the SIAs of the $CrTe_2/In_2Te_3(\uparrow)$ and the $CrTe_2/In_2Te_3(\downarrow)$ prefer in-plane and out-of-plane magnetic easy axes, respectively. Taking them together, we obtain that the FM Heisenberg exchange couplings dominate over the DMIs and thus give rise to a FM magnetic ground state with its magnetization determined by the SIA,[35] consistent with our MC simulated results.

Figure 3a shows the Γ-dependent Gilbert dampings of $CrTe_2/In_2Te_3$ with upward and downward FE polarizations of $In_2Te_3$. Similar to previous studies,[36,37] the Gilbert dampings of $CrTe_2/In_2Te_3$ decrease first and then increase as the scattering rate Γ increases. Astonishingly, the Gilbert dampings of $CrTe_2/In_2Te_3(\uparrow)$ and $CrTe_2/In_2Te_3(\downarrow)$ are distinctly different at the same scattering rate Γ ranging from 0.001 to 1.0 eV. To have a more intuitive sense on the effect of the FE polarizations of $In_2Te_3$ on the Gilbert dampings in $CrTe_2/In_2Te_3$, we calculate the ratio $\eta = \alpha_\uparrow/\alpha_\downarrow$ at any given Γ, where $\alpha_\uparrow(\alpha_\downarrow)$ is the Gilbert damping of CrTe$_2$/In$_2$Te$_3$(↑) [CrTe$_2$/In$_2$Te$_3$(↓)]. As shown in Fig. 3b, the ratio $\eta$ ranges from 6 to around 1.3 with increasing Γ. As the FE polarization of In$_2$Te$_3$ can be switched from upward to downward by an external electric field, the Gilbert damping of CrTe$_2$/In$_2$Te$_3$ is electrically tunable in practice.

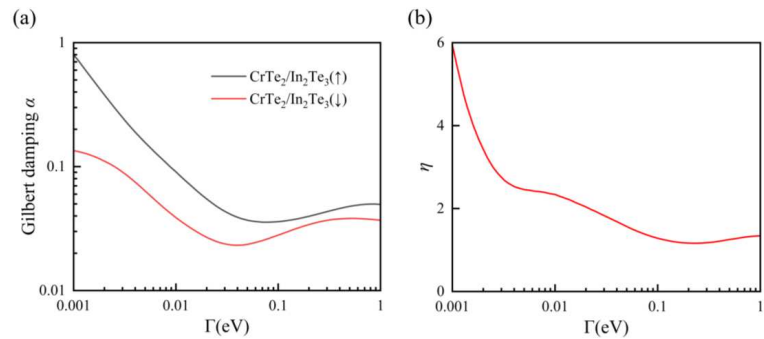

FIG. 3. (a) The Γ-dependent Gilbert dampings of CrTe$_2$/In$_2$Te$_3$ with upward (black line) and downward (red line) FE polarizations of In$_2$Te$_3$. (b) The Gilbert damping ratio $\eta$ as a function of the scattering rate Γ.

To gain a deep insight into how the FE polarization of In$_2$Te$_3$ tunes the Gilbert damping in CrTe$_2$/In$_2$Te$_3$, we investigate the *k*-dependent contributions to the Gilbert dampings of CrTe$_2$/In$_2$Te$_3$(↑) and CrTe$_2$/In$_2$Te$_3$(↓). As shown in Fig. 4a and 4b, the bands around Fermi level have qiute different intermixing between CrTe$_2$ and In$_2$Te$_3$ states when the FE polarizaiton of In$_2$Te$_3$ is reversed. Explicitly, there are obvious intermixing below Fermi leve in CrTe$_2$/In$_2$Te$_3$(↑) while the intermixing mainly takes place above Fermi level in CrTe$_2$/In$_2$Te$_3$(↓). Especially, the bands intersected by Fermi level are at different *k* points in CrTe$_2$/In$_2$Te$_3$(↑) and CrTe$_2$/In$_2$Te$_3$(↓). Through looking into the *k*-dependent contributions to their Gilbert dampings (Fig. 4c and 4d), we see that large contributions are from the *k* points (highlighted by arrows in Fig. 4) at which the bands of CrTe$_2$ cross Fermi level. In addition, these large contributions are different. Such *k*-dipendent contribution to Gilbert dampings is understandable. Based on the scattering theory of Gilbert damping [38], Gilbert damping parameter is calculated using the following Eq. (3) [36]

$$\alpha_{\mu\nu} = -\frac{\pi\hbar\gamma}{M_S}\sum_k\sum_{ij}\langle\psi_{k,i}|\frac{\partial H_k}{\partial u_\mu}|\psi_{k,j}\rangle\langle\psi_{k,j}|\frac{\partial H_k}{\partial u_\nu}|\psi_{k,i}\rangle\delta(E_F-E_{k,i})\delta(E_F-E_{k,j}) \quad (3),$$

where $E_F$ is Fermi level and $E_{k,i}$ is the enery of band $i$ at a given $k$ point. Due to the delta $\delta(E_F-E_{k,i})\delta(E_F-E_{k,j})$, only the valence and conduction bands near Fermi level make dominant contribution to the Gilbert damping. Additionally, their contributions also depend on factor $\langle\psi_{k,i}|\frac{\partial H_k}{\partial u_\mu}|\psi_{k,j}\rangle\langle\psi_{k,j}|\frac{\partial H_k}{\partial u_\nu}|\psi_{k,i}\rangle$. Overall, through changing the intersections between the bands of CrTe$_2$ and Fermi level, the reversal of the FE polarization of In$_2$Te$_3$ can modulate the contributions to Gilbert damping. Consequently, the total Gilbert dampings are different in CrTe$_2$/In$_2$Te$_3$(↑) and CrTe$_2$/In$_2$Te$_3$(↓).

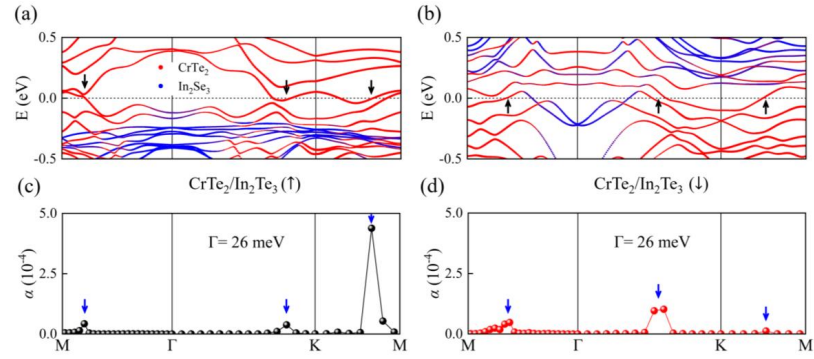

FIG. 4. (a) Band structure calculated with spin-orbit coupling and (c) the $k$-dependent contributions to the Gilbert damping in CrTe$_2$/In$_2$Te$_3$(↑). (b) and (d) same as (a) and (c) but for CrTe$_2$/In$_2$Te$_3$(↓). In (a) and (b), Fermi levels are indicated by horizontal dash lines and the states from CrTe$_2$ and In$_2$Te$_3$ are shown by red and blue, respectively.

From experimental perspectives, the fabrication of CrTe$_2$/In$_2$Te$_3$ vdW heterostructure should be feasible. On the one hand, CrTe$_2$ with the lattice constant of 3.95 Å has been successfully grown on ZrTe$_2$ substrate by the molecular beam epitaxy.[29] On the other hand, In$_2$Te$_3$ is also synthesized.[39] Taking these and the vdW nature of CrTe$_2$ and In$_2$Te$_3$ together, a practical scheme of growing CrTe$_2$/In$_2$Te$_3$ is sketched in Fig. S8: first grow CrTe$_2$ ML on ZrTe$_2$ substrate[29] and then put In$_2$Te$_3$ ML on CrTe$_2$ to form the desired CrTe$_2$/In$_2$Te$_3$ vdW heterostructure.

In summary, by constructing a vdW heterostructure of 2D FM metal CrTe$_2$ and FE In$_2$Te$_3$ MLs, we find that the magnetic properties of CrTe$_2$ are engineered by the reversal of the FE polariton of In$_2$Te$_3$. Although the ferromagnetism of CrTe$_2$ is maintained in the presence of the FE In$_2$Te$_3$, its magnetic easy axis can be tuned from in-plane to out-of-plane by reversing the FE polarization of In$_2$Te$_3$. More importantly, the Gilbert damping of CrTe$_2$ is tunable with a wide range when reversing the FE polarization of In$_2$Te$_3$ from upward to downward. Such tunability of the Gilbert damping in CrTe$_2$/In$_2$Te$_3$ results from the changed intersections between the bands of CrTe$_2$ and Fermi level on reversing the FE polarizations of In$_2$Te$_3$. Our work introduces a remarkably useful method to electrically tune the Gilbert dampings of 2D vdW FM metals by contacting them with ferroelectrics, and should stimulate more experimental investigations in this realm.

See the supplementary material for the details of computational methods[31,36,40-50] and other results mentioned in the main text.

This project is supported by National Nature Science Foundation of China (No. 12104518, 92165204, 11974432), NKRDPC-2018YFA0306001, NKRDPC-2022YFA1402802, GBABRF-2022A1515012643 and GZABRF-202201011118. Density functional theory calculations are performed at Tianhe-II.

## AUTHOR DECLARATIONS
**Conflict of Interest**

The authors have no conflicts to disclose.

**Author Contributions**

**Liang Qiu**: Investigation (equal); Methodology (equal); Writing –original draft (equal). **Zequan Wang**: Methodology (equal). **Xiao-sheng Ni**: Investigation (equal); Methodology (equal). **Dao-Xin Yao**: Supervision (equal); Funding acquisition (equal); Investigation (equal); Writing – review &editing (equal). **Yusheng Hou**: Conceptualization (equal); Funding acquisition (equal); Investigation (equal); Project administration(equal); Resources (equal); Supervision (equal); Writing – review &editing (equal).

## DATA AVAILABILITY

The data that support the findings of this study are available from the corresponding authors upon reasonable request.

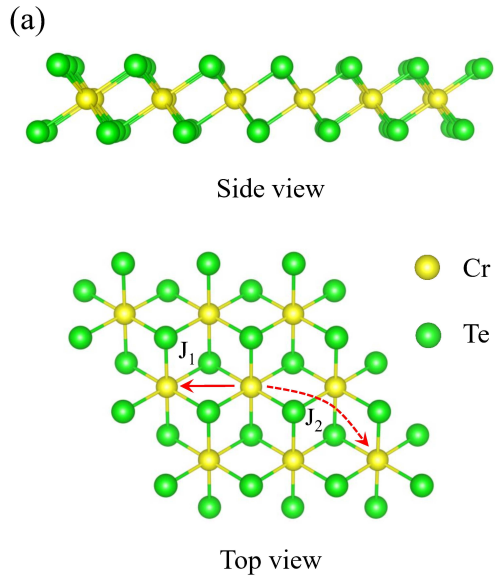
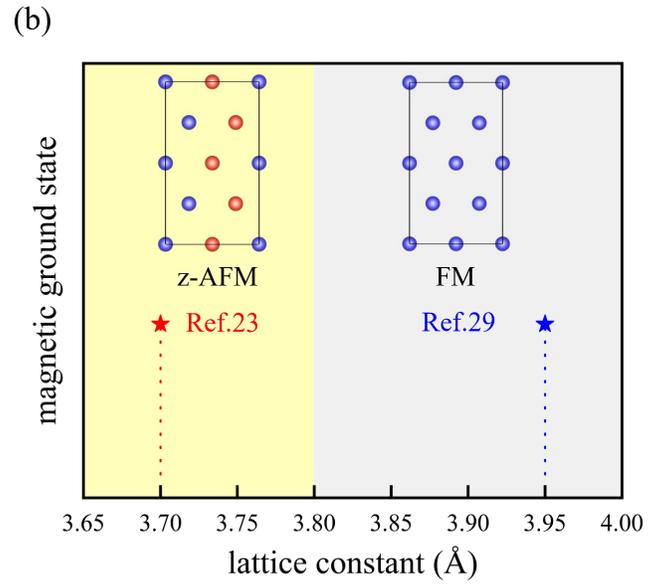

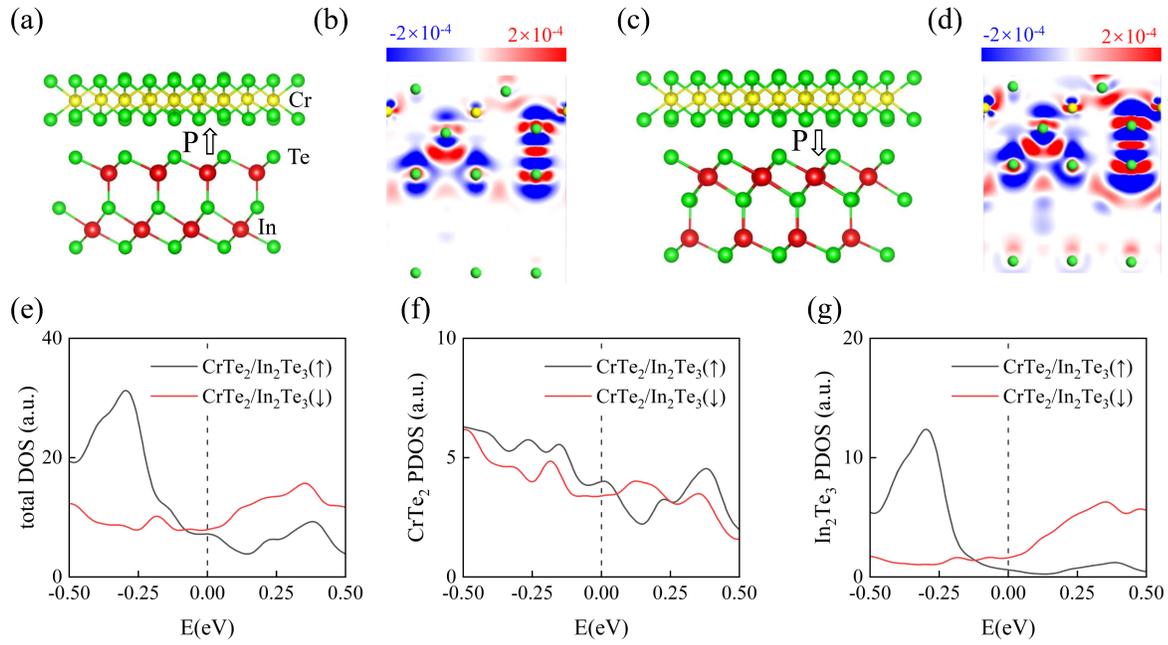

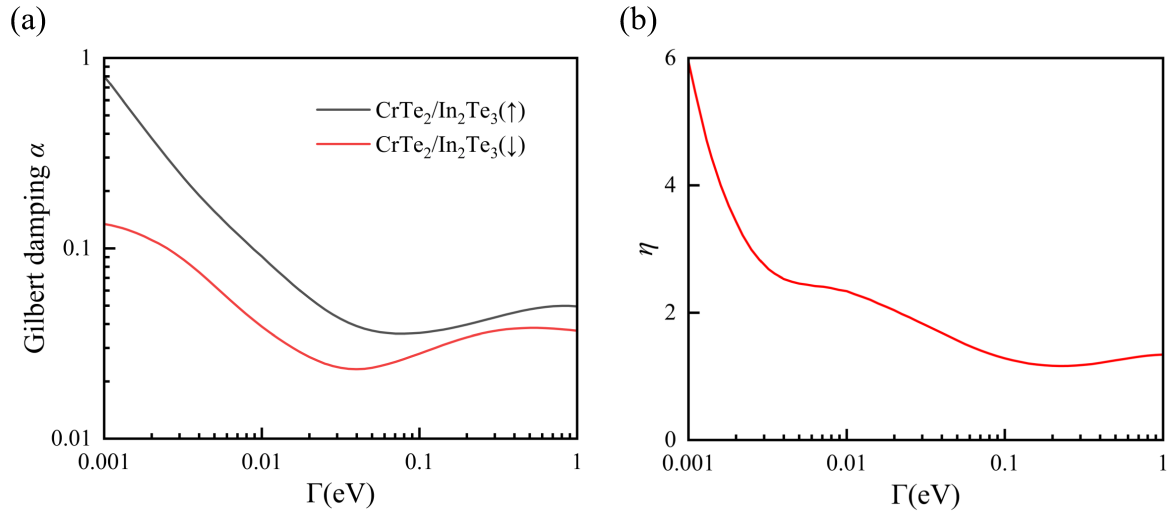

(a) (b)

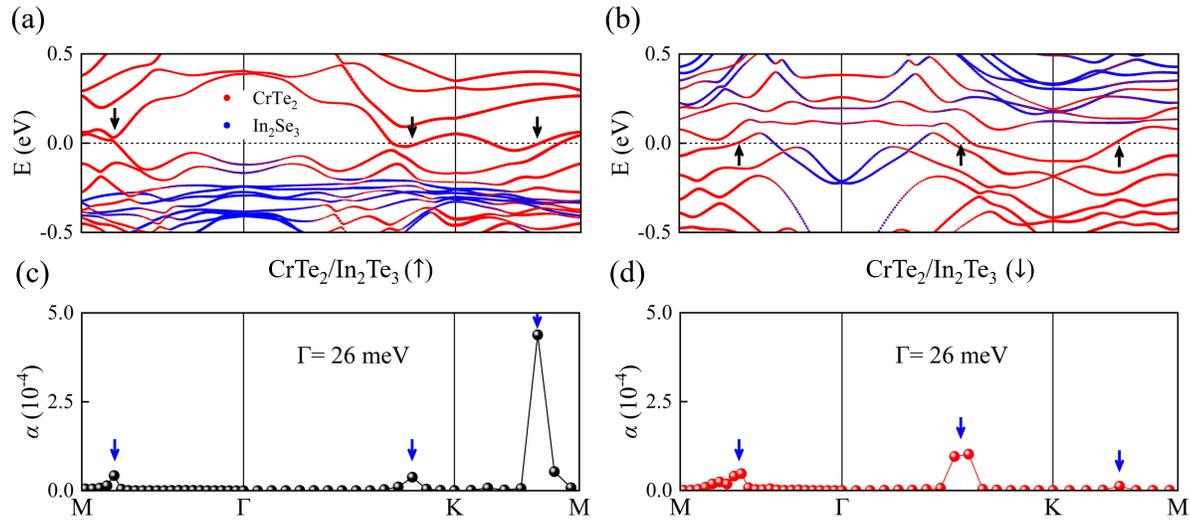